\documentclass{article}
\usepackage[T1]{fontenc}
\usepackage[utf8]{inputenc}
\usepackage[]{ismir}
\usepackage{amsmath,cite,url}
\usepackage{graphicx}
\usepackage{color}

\usepackage{url}
\usepackage{adjustbox}
\usepackage{booktabs} 
\usepackage{pifont}
\usepackage{multirow}
\usepackage{makecell}
\usepackage{amsfonts} 
\usepackage{enumitem} 
\usepackage{gensymb} 
\newcommand{\topic}[1]{\par\vspace{0.2\baselineskip}{\noindent\textbf{#1 --- }}}

\title{
Enhancing Neural Audio Fingerprint Robustness to Audio Degradation for Music Identification
}

\multauthor
  {R. Oguz Araz$^1$ \hspace{1cm} Guillem Cort\`es-Sebasti\`a$^2$ \hspace{1cm} Emilio Molina$^2$ \hspace{1cm} Joan Serr\`a$^3$ }
  {{\bf Xavier Serra$^1$ \hspace{1cm} Yuki Mitsufuji$^{3,4}$ \hspace{1cm} Dmitry Bogdanov$^1$}\\
  $^1$ Music Technology Group, Universitat Pompeu Fabra, Barcelona, Spain\\
  $^2$ BMAT Licensing S.L., Barcelona, Spain\\
  $^3$ Sony AI \hspace{0.1cm} $^4$ Sony Group Corporation\\
  {\tt\small recepoguz.araz@upf.edu}
  }

\def\authorname{R. O. Araz, G. Cort\`es-Sebasti\`a, E. Molina, J. Serr\`a, X. Serra, Y. Mitsufuji, and D. Bogdanov}

\usepackage[bookmarks=false,pdfauthor={\authorname},pdfsubject={\pdfsubject},hidelinks]{hyperref}

\begin{document}

\maketitle

\begin{abstract}
Audio fingerprinting (AFP) allows the identification of unknown audio content by extracting compact representations, termed audio fingerprints, that are designed to remain robust against common audio degradations. Neural AFP methods often employ metric learning, where representation quality is influenced by the nature of the supervision and the utilized loss function. However, recent work unrealistically simulates real-life audio degradation during training, resulting in sub-optimal supervision. Additionally, although several modern metric learning approaches have been proposed, current neural AFP methods continue to rely on the NT‑Xent loss without exploring the recent advances or classical alternatives. In this work, we propose a series of best practices to enhance the self-supervision by leveraging musical signal properties and realistic room acoustics. We then present the first systematic evaluation of various metric learning approaches in the context of AFP, demonstrating that a self‑supervised adaptation of the triplet loss yields superior performance. Our results also reveal that training with multiple positive samples per anchor has critically different effects across loss functions. Our approach is built upon these insights and achieves state-of-the-art performance on both a large, synthetically degraded dataset and a real-world dataset recorded using microphones in diverse music venues.
\end{abstract}

\section{Introduction}\label{sec:introduction}

Audio fingerprinting (AFP) methods are used for identifying unknown audio content~\cite{haitsma_robust_2001, haitsma_highly_2002, cano_review_2002}. Identification is achieved by extracting distinctive representations, known as fingerprints, from audio segments, which are matched against a database of reference audio. Possible AFP applications include music identification (MI)~\cite{wang_shazam_2006}, duplicate detection~\cite{burges_distortion_2003}, and broadcast monitoring~\cite{cortes_baf_2022}. This paper focuses on MI, where scalability is crucial due to large databases, requiring compact fingerprints for efficient retrieval. MI can be performed at both the track-level and segment-level, the latter involving time-aligned matching between query audio and reference tracks.

In MI, query audio can significantly differ from its reference track due to various factors. Intentional transformations, such as pitch shifting and time stretching, can alter a track, and since MI often relies on microphone recordings of playback sounds, the query audio may be degraded due to signal transduction and environmental factors during sound propagation. For accurate identification, fingerprints should be robust enough to be recognized despite such alterations. One existing approach, NowPlaying~\cite{arcas_now_2017}, learns to create robust fingerprints from real recordings of aligned clean-degraded music pairs. However, such datasets are costly to gather and, hence, most methods rely on self-supervision from synthetic audio degradations.~\cite{yu_contrastive_2020, chang_neural_2021, singh_attention-based_2022, bhattacharjee_grafprint_2025}.

The quality of the representations obtained by self-supervised learning (SSL) is influenced by various factors, including the nature of the self-supervision signal and the calculation of the loss function. Yet, current neural AFP approaches unrealistically simulate real-life audio degradation during training, providing sub-optimal supervision. Moreover, although recent work in neural AFP generally assumes that the NT‑Xent~\cite{chen_simple_2020} loss outperforms other classical losses on MI, a systematic comparison between these objectives is missing. Besides, several modern contrastive self‑supervised losses that build on the NT-Xent loss have shown promise in related audio tasks~\cite{serra_supervised_2025, guinot_semi-supervised_2024}, but have not been evaluated in the context of AFP. 
In addition, some of these loss functions allow the construction of training batches with multiple differently degraded versions of the same clean audio, 
a strategy that could improve identification but remains unexplored within the scope of AFP.

Our work builds upon NAFP~\cite{chang_neural_2021}, a neural AFP approach that generates real-valued fingerprints, with open-source code and pre-trained weights. We uncover several sub-optimal design choices of NAFP regarding its SSL strategy and treatment of room acoustics. To address these issues, we propose a series of best practices that significantly enhance identification performance. Moreover, we reveal critical implementation problems in NAFP’s evaluation method that skew performance metrics. We revise these problems and conduct evaluations on both synthetically degraded queries and a real-world dataset recorded with microphones across diverse music venues. Next, we assess the effectiveness of different loss functions for AFP, demonstrating that the triplet loss, contrary to common belief, delivers the best performance. We then explore the benefits and drawbacks of increasing the number of positive samples per anchor in a batch, providing practical training guidelines. Building on these improvements, we obtain a state-of-the-art approach, outperforming the publicly available baselines. We share our pre-trained model weights, open-source code, and data.\footnote{\url{https://github.com/raraz15/neural-music-fp}}

\section{Related work}\label{sec:related}

\subsection{Metric learning}\label{sec:related:deep}

Metric learning aims to create representations in which semantically similar elements are closer to each other than semantically dissimilar ones. 
The triplet loss~\cite{schroff_facenet_2015} is a supervised metric learning function that is proven effective in various retrieval tasks. It uses class labels to form triplets of anchor, positive, and negative samples. 

Recent work leverages self-supervised metric learning, with NT-Xent being a particularly successful loss function~\cite{chen_simple_2020}. More recent work then builds upon NT-Xent: DCL~\cite{yeh_decoupled_2022} decouples positive pair's similarities from the negative pairs'; A\&U~\cite{wang_understanding_2020} proposes two qualities that good representations should obtain, and theoretically proves that using NT-Xent with an infinite-size batch optimizes them; KCL~\cite{koromilas_bridging_2024} re-formulates A\&U to require a smaller batch.

NT-Xent and its extensions are only defined for a single positive sample per anchor. SupCon~\cite{khosla_supervised_2020} (supervised) and MultiPosCon~\cite{tian_stablerep_2023} (self-supervised) both extend NT-Xent to multiple positives per anchor, reporting benefits in image-related tasks. Notably, MultiPosCon and NT-Xent are equivalent when each anchor has only one positive. Hence, for simplicity, we treat them synonymously. 

The quality of learned representations is influenced by how the batches are constructed, particularly by the number of anchor samples and the number of positives per anchor~\cite{schroff_facenet_2015}. However, these aspects remain unexplored in AFP. To address this gap, we perform controlled experiments. Throughout this paper, let $N_{\text{A}}$ denote the number of anchors in a batch and $N_{\text{PPA}}$ the number of positives per anchor. The number of total samples in a batch is then $N_{\text{B}}=N_{\text{A}}\left(1+N_{\text{PPA}}\right)$.

\subsection{Neural AFP}\label{sec:related:neuralaudio}

Neural AFP approaches can be broadly classified into two categories based on the generated fingerprints: binary and real-valued. In this work, we focus on the latter. Relevant approaches include NowPlaying~\cite{arcas_now_2017}, which employs the triplet loss to train a convolutional neural network (CNN), and also CULAF~\cite{yu_contrastive_2020}, NAFP~\cite{chang_neural_2021}, and ABFP~\cite{singh_attention-based_2022}, all of which use the NT-Xent loss to train CNNs. Additionally, GraFP~\cite{bhattacharjee_grafprint_2025} trains a graph CNN with the NT-Xent loss.

Table~\ref{tab:approaches} compares these approaches in terms of two aspects that determine an AFP system's use case: audio sampling rate and fingerprint generation rate (measured by the hop duration). We do not consider high sampling rates necessary for MI, as query audio is often transmitted over bandwidth-limited channels. Moreover, since humans can typically identify music even under low-pass filtering, a robust system should be capable of doing the same. Likewise, the hop sizes around 100\,ms used by GraFP and ABFP substantially increase the time, storage, and computational costs of fingerprint extraction and retrieval, reducing overall scalability. This becomes evident by looking at their reported test databases containing 30-second audio chunks instead of full tracks, selected from the Free Music Archive (FMA)~\cite{defferrard_fma_2017}. They contain about 97\,k (\texttt{fma\_large}) and 25\,k (\texttt{fma\_medium}) chunks, corresponding to 28.3\,M and 7.3\,M fingerprints, respectively. By contrast, NAFP's test database (and ours, as discussed in Section~\ref{sec:method:eval}) contains over 93\,k full-length tracks (\texttt{fma\_full}), totaling 53.6\,M fingerprints. Using a 100\,ms hop duration would result in 268\,M fingerprints, requiring much longer time for both extraction and retrieval.

\begin{table}[t]
    \centering
    \begin{adjustbox}{max width=\columnwidth}
    \begin{tabular}{lcccc}
    \toprule
        Model & $F_\text{S}$ [Hz] & $T_\text{L}$ [s] & $T_\text{H}$ [s] & Available \\
    \midrule
        NowPlaying \cite{arcas_now_2017} & n/a & n/a & 1.000 & \ding{55} \\
        CULAF \cite{yu_contrastive_2020} & 16\,k & 2.50 & 2.125 & \ding{55} \\
        NAFP \cite{chang_neural_2021} & ~~8\,k & 1.00 & 0.500 & \ding{51} \\
        ABFP \cite{singh_attention-based_2022} & 16\,k & 0.96 & 0.096 & \ding{55} \\
        GraFP \cite{bhattacharjee_grafprint_2025} & 16\,k & 1.00 & 0.100 & \ding{51} \\
        NMFP (proposed) & ~~8\,k & 1.00 & 0.500 & \ding{51} \\
    \bottomrule
    \end{tabular}
    \end{adjustbox}
    \caption{
    Real-valued neural AFP models. Columns indicate the sampling rate ($F_\text{S}$), segment duration ($T_\text{L}$), hop duration ($T_\text{H}$), and publicly available code and weights.
    }
    \label{tab:approaches}
\end{table}

\section{Methodology}\label{sec:method}

In this section, we outline NAFP’s methodology and, where applicable, describe our modifications.

\subsection{Audio degradations}\label{sec:method:degradation}

NAFP authors focus on achieving robustness against three types of audio degradation: additive background noise, room reverberation, and microphone response. We follow their degradation chain, but curate more extensive sets for each type. NAFP uses background noise recordings featuring a mix of random noises and two specific acoustic scenes: subway and pub environments. However, we found this selection to lack sufficient diversity, and instead adopted the TUT Acoustic Scenes 2016 dataset~\cite{mesaros_tut_2016}, which includes 15 distinct acoustic scenes representing various potential MI use cases. The training and test sets consist of 585 and 195 minutes of recordings, respectively.

For room impulse responses (IRs), we use OpenAIR~\cite{murphy_openair_2010} and AIR~\cite{jeub_binaural_2009} datasets, similar to NAFP, but with adjustments considering room acoustics. From OpenAIR, we use all 143 mono and stereo recordings from 28 diverse environments, such as halls and churches, chosen for their long-duration IRs. From AIR, we use the IRs recorded without a dummy head. For the binaural recordings, we include both channels separately, as they capture different reflection patterns, resulting in 60 IR measurements from 6 rooms. Unlike NAFP, we also utilize the MIT-Survey~\cite{traer_statistics_2016} dataset, contributing 270 IRs from various public spaces.

For microphone IRs, we exclusively use the dataset provided by Franco et al. \cite{franco_multi-angle_2022}. It contains measurements of 25 microphones encompassing 38 unique microphone-polar pattern combinations. The IRs were measured at distances of 0.5\,m, 1.25\,m, and 5\,m and at multiple incident angles, from which we chose integer multiples of 60\degree.

We partition the recordings for each degradation into train and test sets. For background noise, we use the partitions proposed in Mesaros et al.~\cite{mesaros_tut_2016}. For IRs, we ensure that the measurements of the same room or microphone are contained in either the training or test set. This consideration does not hold in the publicly available IR partitions of NAFP, which can be verified by comparing the filenames.

\subsection{Training}\label{sec:method:train}

NAFP authors experiment with two variations based on the training optimizer: Adam~\cite{kingma_adam_2015} and LAMB~\cite{you_large_2020}. The authors reported that LAMB requires a much larger batch size to achieve high performance, making training on consumer-grade GPUs impractical. As a result, we adopt the Adam variation as our baseline, whose pre-trained weights are unavailable. Therefore, we train NAFP-Adam ourselves using the official implementation and the provided audio degradation datasets.\footnote{\url{https://github.com/mimbres/neural-audio-fp/}} In all our experiments, we use the same training music as NAFP: 10\,k~audio chunks of 30\,s duration, sourced from \texttt{fma\_medium}.

NAFP 
has an audio context of one second, 
and fingerprints are extracted with a 500\,ms hop duration at inference. At training, a $\pm$\,200\,ms random offset is applied between positive and anchor samples to enhance robustness to potential misalignment between query and reference fingerprints. We increase this offset to $\pm$\,250\,ms, corresponding to 50\% of the hop duration, which we find more intuitive.
As for input features, we follow the same magnitude mel-spectrogram extraction parameters as NAFP but apply a slightly different formulation during the power conversions step. Finally, we scale the features to $[-1, 1]$, using the global dynamic range of $80$\,dB.

We train all models for 100~epochs with the same architecture, optimizer, learning rate scheduler, SpecAugment~\cite{park_specaugment_2019} implementation, and NT-Xent parameter $\tau$ as in NAFP-Adam. However, we train our models using automatic mixed precision. The configuration $N_{\text{A}}=768$, $N_{\text{PPA}}=1$ requires approximately 15 hours to complete on a single NVIDIA RTX 4090 GPU and 20 CPU cores.

\subsection{Retrieval}\label{sec:method:retrieval}
NAFP performs a two-stage retrieval using Faiss~\cite{johnson_billion-scale_2019}, a library for efficient large-scale similarity search. 
First, for each fingerprint in the query sequence, the top 20 approximately most similar segments are retrieved. Then, a candidate sequence is constructed for each unique segment, considering its position in the query sequence, and the average similarity score is calculated between the query and candidate sequences. The original Faiss index type and hyper-parameters are kept unchanged in our work.

\subsection{Evaluation}\label{sec:method:eval}

NAFP measures identification performance using the Top-1 hit rate metric, which we also employ. However, whereas NAFP evaluates only segment-level identification (focusing on exact and near matches within one hop duration), we evaluate both track- and segment-level identification.

We discovered that, in NAFP's evaluation, some query tracks were represented up to 11~times, while others were not represented at all, skewing the results. To address this, we 
use 
30-second 
audio chunks 
as queries 
and select 6 equally spaced indices within the chunk. Starting from each index, we query fingerprint sequences of lengths 1, 3, 9, and 19, corresponding to 1, 2, 5, and 10~seconds of audio, respectively. This method efficiently utilizes the chunks and ensures that each track contributes uniformly to the metric. Additionally, NAFP's fingerprint storage implementation simply concatenates fingerprints from all query chunks, disregarding track boundaries. As a result, 12.7\% of the query sequences contain fingerprints from two tracks. In our implementation, we ensure that each query sequence is contained within a single track.

Various approaches, including NAFP, perform evaluations on synthetically degraded queries~\cite{haitsma_robust_2001, haitsma_highly_2002, sonnleitner_robust_2016, baez-suarez_samaf_2020, agarwaal_robust_2023}, while only a few consider real microphone recordings~\cite{arcas_now_2017, ramona_audioprint_2013}. We evaluate our models on both types of data.

\topic{Syntheticly degraded queries}
We build our synthetic evaluation on NAFP's publicly available test set, which is a subset of \texttt{fma\_full}. However, it is known that FMA contains duplicates, so we perform duplicate removal. Their test database contains 93,358 full-length tracks. We noted that 1,205 unique tracks were not included, which we included in our test database, resulting in 95,163 total tracks, distinct from the training tracks.

For the query set, NAFP uses audio chunks of 30-second duration from 500 tracks, which is insufficient for a comprehensive evaluation. A larger set increases statistical power, offers better insight into robustness across various degradations, and reduces potential biases. 
Therefore, we additionally incorporate 9,500 random tracks from our test database, bringing the total to 10,000 query tracks. 
We then degrade each track from start to end by sequentially applying additive background noise (using a random SNR sampled uniformly from [0,10]~dB), followed by convolution with room and microphone IRs (interleaved with random gain, using full IR durations). From the resulting audio, we randomly sample 30-second chunks to be used as queries. We make this dataset publicly available.

It is worth emphasizing that our evaluation setup uses significantly more queries than those used in NAFP, GraFP, and ABFP, and our database is substantially larger than those used in GraFP and ABFP. Therefore, our evaluation provides a more comprehensive and realistic assessment of scalability than prior work. We argue that scalability claims should be evaluated under similar conditions.


\begin{table*}[t]
    \renewcommand{\arraystretch}{0.85}
    \setlength{\tabcolsep}{8pt}
    \centering
    \begin{adjustbox}{max width=\linewidth}
    \begin{tabular}{cccccccccccccc}
    \toprule
        \multirow{3}{*}{\vspace{-12pt}Row}& \multirow{3}{*}{\vspace{-12pt}Approach} & \multirow{3}{*}{\vspace{-12pt}FN} & \multirow{3}{*}{\vspace{-12pt}$T_{\text{IR}}$} & \multirow{3}{*}{\vspace{-12pt}Past} & \multirow{3}{*}{\vspace{-12pt}$F_{\text{LC}}$} & \multicolumn{8}{c}{Top-1 hit rate (\%)} \\
        \cmidrule(lr){7-14}
         & & & & & & \multicolumn{4}{c}{Synthetic data} & \multicolumn{4}{c}{Real-world data} \\
        \cmidrule(lr){7-10}\cmidrule(lr){11-14}
         & & & & & & 1\,s & 2\,s & 5\,s & 10\,s & 1\,s & 2\,s & 5\,s & 10\,s \\
    \midrule
        1 & Baseline & \ding{55} & 75\,ms & \ding{55} & 300\,Hz & 72.2 & 82.8 & 89.7 & 92.3 & 33.6 & 49.9 & 65.4 & 71.7 \\
    \midrule
        2 & \multirow{5}{*}{Ours} & \ding{55} & 75\,ms & \ding{55} & 300\,Hz & 74.4 & 83.8 & 89.9 & 92.1 & 39.4 & 56.2 & 70.5 & 76.1 \\
        3 & & \ding{51} & 75\,ms & \ding{55} & 300\,Hz & 76.7 & 85.8 & 91.6 & 93.6 & 45.5 & 64.8 & 77.3 & 81.9 \\
        4 & & \ding{51} & 1\,s & \ding{55} & 300\,Hz & 77.9 & 86.7 & 92.0 & 93.8 & 49.3 & 66.7 & 78.5 & 82.6 \\
        5 & & \ding{51} & 1\,s & \ding{51} & 300\,Hz & 80.1 & 87.9 & 92.5 & 94.2 & 50.6 & 68.1 & 79.2 & 83.0 \\
        6 & & \ding{51} & 1\,s & \ding{51} & 160\,Hz & \textbf{80.5} & \textbf{88.0} & \textbf{92.6} & \textbf{94.2} & \textbf{54.5} & \textbf{70.7} & \textbf{81.0} & \textbf{84.3} \\
    \bottomrule
    \end{tabular}
    \end{adjustbox}
    \caption{
    Improvements over NAFP in track-level identification using our test set. `FN' indicates if the false negatives issue in the batches is corrected, $T_{\text{IR}}$ denotes the impulse response truncation duration, `Past' indicates whether the acoustic history is applied during reverberation, and $F_{\text{LC}}$ specifies the lower cut-off frequency during feature extraction.
    }
    \label{tab:improvements}
\end{table*}

\topic{Real-world data}
We conduct a real-world evaluation in collaboration with BMAT Licensing S.L. using a dataset of microphone recordings captured using smartphones and digital recorders in various settings, including bars, nightclubs, and concert halls. The ground truth was established by a fingerprinting system against a large database, with results verified by human annotators. All reference tracks have at least one match, totaling 3,692 query-reference pairs. Although the dataset is not large enough to test scalability, it serves well for testing robustness in real environments.

\section{Neural Music Fingerprinting}\label{sec:nmfp}

We propose a new approach for training neural AFP models that focuses on musical signal properties and real-life audio degradation. First, to establish a baseline, we benchmark the NAFP-Adam model on track-level identification. Next, we incrementally apply a series of best practices to this baseline to facilitate the learning process, validating the contribution of each. Finally, we explore a range of metric learning approaches to further enhance performance. We refer to our method as NMFP (Neural Music Fingerprinting) to highlight its focus on MI.

\subsection{Best practices for MI}\label{sec:nmfp:improvements}

In this section, we improve the self-supervision during training by more closely simulating real-life audio degradation, eliminating faulty learning signals, and providing the model with additional cues. Altogether, our practices improve identification performance by 8.3\% and 20.9\% on the synthetic and real-world datasets, respectively (Table~\ref{tab:improvements}). Notably, our practices add negligible training overhead while substantially raising performance, making them our recommended best practices for training.

\topic{Appropriate audio degradation datasets}
Self-supervision comes from the objective of embedding the clean audio and its synthetically degraded version close to each other. By exposing the model to degradations that closely resemble those encountered in real-world scenarios, we encourage structuring the learned representations accordingly. Therefore, the choice of degradation data is critical; it should reflect realistic use cases. Training with our degradation datasets as opposed to those of NAFP increases performance on both datasets (Table~\ref{tab:improvements}, rows 1--2).

\topic{Eliminating false negatives} 
The objective of the NT-Xent loss is to increase the similarity between an anchor and its positive, while decreasing the similarity with all its negatives in the batch. However, in NAFP, a batch can contain multiple anchor samples from the same track with an 18\% probability (64 segments chosen from 590,000 segments across 10,000 tracks with 59 segments each). Belonging to the same track, these samples share various musical properties. Yet, the loss function treats them as negative pairs, resulting in a faulty learning signal. In our implementation, we eliminate false negatives by ensuring that each batch contains one anchor sample per track. This increases fingerprint distinctiveness (Table~\ref{tab:improvements}, rows 2--3).

\topic{Full IRs} 
During training, NAFP truncates all IRs to 75~ms. For a room IR, this duration only includes the early reflections~\cite{beranek_concert_1992}. We aim for more realistic degradation so that our models can learn to generate robust fingerprints against real-life reverberation. Therefore, we use the full duration of IRs, which can go up to several seconds for large rooms \cite{traer_statistics_2016}. However, in practice, we truncate IRs to the model's context length, as the longer part will not contribute to the current segment. The resulting model exhibits improved robustness (Table~\ref{tab:improvements}, rows 3--4).

\topic{Past reverberation degradation} 
Query fingerprints extracted from microphone recordings contain reverberation, including the tails of past sound events. Previous audio degradation methods often overlook this aspect, convolving an audio segment with a room IR as if the sound starts abruptly, without acoustic history. Misrepresenting reverberation can result in learning unrealistic representations, especially in such fine-grained applications.
Hence, we convolve audio segments starting from their past context and discard the past after convolution, yielding a segment that contains the reverberation of current and past events. Matching the model’s context, we use a one-second past context duration. This increases robustness considerably (Table~\ref{tab:improvements}, rows 4--5).

\topic{Lower frequencies} The 300\,Hz lower frequency cutoff applied in NAFP's feature extraction discards valuable information that can provide additional musical cues. In music venues such as concert halls and festival areas, at long distances from the speakers, the bass frequencies will form the majority of the sounds surviving the background noise from the crowd. Since most microphones, including smartphone microphones, can capture lower frequencies, extending the frequency range can provide benefits across different recording devices. We tested multiple values and selected a 160\,Hz bound. The resulting model achieves further improved performance (Table~\ref{tab:improvements}, rows 5--6).

\subsection{Exploring metric learning}\label{sec:nmfp:metric}

Having improved the self-supervision, we now explore different metric learning methods to further enhance the identification performance. This exploration includes comparing several loss functions, investigating the effect of training with different numbers of anchors and positives per anchor, and tuning loss function hyper-parameters.

\begin{table}[t]
    \centering
    \renewcommand{\arraystretch}{0.85}
    \setlength{\tabcolsep}{11pt}
    \begin{tabular}{lcccc}
    \toprule
        \multirow{3}{*}{\vspace{-12pt}Loss} & \multicolumn{4}{c}{Top-1 hit rate (\%)} \\
        \cmidrule(lr){2-5}
        & \multicolumn{2}{c}{Synthetic data} & \multicolumn{2}{c}{Real-world data} \\
        \cmidrule(lr){2-3}\cmidrule(lr){4-5}
         & 1\,s & 2\,s & 1\,s & 2\,s \\
    \midrule
        NT-Xent & 84.1 & 90.1 & 58.7 & 72.3 \\
            DCL & 83.0 & 89.6 & 54.5 & 68.7 \\
           A\&U & 76.4 & 86.2 & 48.0 & 66.3 \\
            KCL & 38.8 & 60.6 & 18.2 & 41.1 \\
        Triplet & \textbf{86.4} & \textbf{91.6} & \textbf{63.4} & \textbf{75.1} \\
    \bottomrule
    \end{tabular}
    \caption{
    Loss function comparison on track-level identification using $N_{\text{A}} = 768$ and $N_{\text{PPA}} = 1$.
    }
    \label{tab:modern}
\end{table}

\topic{Loss function comparison}
Most neural AFP models rely on NT-Xent without comparison with other losses under consistent settings. Here, we systematically compare multiple losses, searching for additional benefits. Specifically, we consider the triplet, NT-Xent, DCL, KCL, and A\&U losses. For DCL, we use the same $\tau$ parameter as NT-Xent. For A\&U and KCL, we take the default parameters in the respective publications. For the triplet loss, we employ hard positive and semi-hard negative mining, computing the loss function using the squared Euclidean distance with a margin of $\alpha=0.5$. In this experiment, for a fair comparison, we use $N_{\text{A}}=768$ for all losses and use one positive per anchor for the triplet loss ($N_{\text{PPA}}=1$), since the remaining losses are only defined for this setting.

\begin{table}[t]
    \centering
    \renewcommand{\arraystretch}{0.85}
    \setlength{\tabcolsep}{4pt}
    \begin{tabular}{lrccccc}
    \toprule
        \multirow{3}{*}{\vspace{-12pt}Loss} & \multirow{3}{*}{\vspace{-12pt}$N_{\text{A}}$} & \multirow{3}{*}{\vspace{-12pt}$N_{\text{PPA}}$} & \multicolumn{4}{c}{Top-1 hit rate (\%)} \\
        \cmidrule(lr){4-7}
         & & & \multicolumn{2}{c}{Synthetic data} & \multicolumn{2}{c}{Real-world data} \\
        \cmidrule(lr){4-5}\cmidrule(lr){6-7}
         & & & 1\,s & 2\,s & 1\,s & 2\,s \\
    \midrule
        \multirow{5}{*}{NT-Xent}
        & 64 & 1 & 80.5 & 88.0 & 54.5 & 70.7 \\
        & 512 & 1 & 84.1 & 90.2 & 58.2 & 71.8 \\
        & 768 & 1 & 84.1 & 90.1 & 58.7 & 72.3 \\
        & 512 & 2 & 76.9 & 85.5 & 46.0 & 63.4 \\
        & 384 & 3 & 74.6 & 83.7 & 44.2 & 62.0 \\
    \midrule
        \multirow{5}{*}{Triplet}
        & 64 & 1 & 82.8 & 89.4 & 57.4 & 72.1 \\
        & 512 & 1 & 86.1 & 91.4 & 62.5 & 74.7\\
        & 768 & 1 & 86.4 & 91.6 & 63.4 & \textbf{75.1} \\
        & 512 & 2 & \textbf{86.6} & \textbf{91.7} & \textbf{63.9} & 75.0 \\
        & 384 & 3 & 86.1 & 91.2 & 63.0 & 74.6 \\
    \bottomrule
    \end{tabular}
    \caption{
    Effect of $N_{\text{A}}$ and $N_{\text{PPA}}$ for NT-Xent and triplet losses.
    }
    \label{tab:batch}
\end{table}

Table~\ref{tab:modern} reports the results, where the triplet loss outperforms all other loss functions. Compared to its closest competitor, NT-Xent, it scores 2.3\% and 4.7\% higher on the synthetic and real-world data, respectively. Notably, the NT-Xent loss outperforms its extensions: DCL, A\&U, and KCL. While we find the decoupling idea of DCL intuitive, our results do not show an improvement. Based on these results, we retain only the triplet and NT-Xent losses for the remainder of our experiments.

\begin{table}[t]
    \centering
    \renewcommand{\arraystretch}{0.85}
    \setlength{\tabcolsep}{8pt}
    \begin{tabular}{lccccc}
    \toprule
        \multirow{3}{*}{\vspace{-12pt}Loss} & \multirow{3}{*}{\vspace{-12pt}$\tau,\alpha$} & \multicolumn{4}{c}{Top-1 hit rate (\%)} \\
        \cmidrule(lr){3-6}
         & & \multicolumn{2}{c}{Synthetic data} & \multicolumn{2}{c}{Real-world data} \\
        \cmidrule(lr){3-4}\cmidrule(lr){5-6}
         & & 1\,s & 2\,s & 1\,s & 2\,s \\
    \midrule
        \multirow{4}{*}{\makecell[c]{NT-Xent \\ ($\tau$)}}
         & 0.01 & 83.2 & 89.8 & 61.2 & 74.3 \\
         & 0.02 & 83.8 & 90.0 & 61.8 & 74.6 \\
         & 0.05 & 84.1 & 90.1 & 58.7 & 72.3 \\
         & 0.07 & 81.8 & 88.8 & 53.8 & 69.2 \\
    \midrule
        \multirow{3}{*}{\makecell[c]{Triplet \\ ($\alpha$)}}
         & 0.3 & 85.9 & 91.2 & 63.8 & 74.9 \\
         & 0.5 & 86.6 & \textbf{91.7} & \textbf{63.9} & \textbf{75.0} \\
         & 0.7 & \textbf{86.7} & 91.6 & 63.3 & 74.7 \\
    \bottomrule
    \end{tabular}
    \caption{
    Hyper-parameter tuning results on track-level identification for NT-Xent~($N_{\text{A}} = 768, N_{\text{PPA}}=1$) and triplet~($N_{\text{A}} = 512, N_{\text{PPA}}=2$) losses.
    }
    \label{tab:tau-alpha}
\end{table}

\topic{Increasing the number of anchors}
For the NT-Xent loss in Table~\ref{tab:batch}, increasing $N_{\text{A}}$ from 64 to 512 yields a 3.6\% improvement on synthetic data, whereas further increasing to 768 causes a saturation. 
On real-world data, however, increasing $N_{\text{A}}$ from 64 to 768 consistently improves performance. For the triplet loss, training with larger $N_{\text{A}}$ progressively increases performance on both datasets.

\topic{Number of anchors vs positives per anchor}
On the one hand, exposing the model to a diverse set of tracks within a batch is beneficial for learning discriminative representations. On the other hand, presenting multiple degraded versions of the same audio segment can help the model learn invariance to real-world distortions. However, due to the GPU memory constraint, increasing the number of positives per anchor reduces the number of anchors that can fit in a batch, creating a trade-off between diversity and invariance. To create multiple positives for an anchor, we randomly shift the anchor independently and use a different combination of degradations.

\begin{table*}[t]
    \centering
    \renewcommand{\arraystretch}{0.85}
    \setlength{\tabcolsep}{12pt}
    \begin{tabular}{lcccccccc}
    \toprule
        \multirow{3}{*}{\vspace{-12pt}Model} & \multicolumn{8}{c}{Top-1 hit rate (\%)} \\
        \cmidrule(lr){2-9}
         & \multicolumn{4}{c}{Synthetic data} & \multicolumn{4}{c}{Real-world data} \\
        \cmidrule(lr){2-5}\cmidrule(lr){6-9}
         & 1\,s & 2\,s & 5\,s & 10\,s & 1\,s & 2\,s & 5\,s & 10\,s \\
    \midrule
        NAFP-Adam & 72.2 & 82.8 & 89.7 & 92.3 & 33.6 & 49.9 & 65.4 & 71.7 \\
        NAFP-LAMB~\cite{chang_neural_2021} & 73.8 & 83.9 & 90.4 & 92.7 & 42.0 & 58.6 & 71.4 & 76.3 \\
    \midrule
        GraFP-500\,ms & 17.0 & 34.8 & 52.1 & 63.6 & 20.0 & 42.5 & 62.0 & 69.4 \\
        GraFP-100\,ms~\cite{bhattacharjee_grafprint_2025} & 19.7 & 53.6 & 72.3 & 80.0 & 46.6 & 60.8 & 74.8 & 82.5 \\
    \midrule
        NMFP (proposed) & \textbf{86.6} & \textbf{91.7} & \textbf{94.5} & \textbf{95.6} & \textbf{63.9} & \textbf{75.0} & \textbf{82.0} & \textbf{84.6} \\
    \bottomrule
    \end{tabular}
    \caption{
    Final comparison on track-level identification.
    }
    \label{tab:bmat}
\end{table*}

In Table~\ref{tab:batch}, when the number of total samples in a batch is set to 1536, switching from $N_{\text{PPA}}=1$ to $N_{\text{PPA}}=2$ or $N_{\text{PPA}}=3$ significantly degrades NT-Xent's performance on both datasets. Moreover, when $N_{\text{A}}=512$, using $N_{\text{PPA}}=2$ performs significantly worse than $N_{\text{PPA}}=1$ on both datasets. These two observations show that, for NT-Xent, using more than one positive per anchor is detrimental, which is not caused by the reduced number of anchors. Indeed, training with $N_{\text{A}} = 64$ and $N_{\text{PPA}}=1$ (roughly ten times smaller batch size) performs significantly better on both datasets. Therefore, we conclude that $N_{\text{PPA}}=1$ is the best setting for NT-Xent. For the triplet loss, Table~\ref{tab:batch} shows that training with $N_{\text{PPA}}=2$ yields the best performance and, unlike NT-Xent, increasing $N_{\text{PPA}}$ does not deteriorate the performance significantly. Notably, training with $N_{\text{PPA}} = 3$ and $N_{\text{PPA}} = 1$ performs comparably, even though the musical variety in a batch is half the amount.

\topic{Hyper-parameter tuning}
In Table~\ref{tab:batch}, we have seen that the triplet loss outperforms NT-Xent in all $N_{\text{PPA}}$-$N_{\text{A}}$ combinations. To gain insight into the effect of hyper-parameters, we experiment with different $\tau$ (NT-Xent) and $\alpha$ (triplet) values. 
The results are given in Table~\ref{tab:tau-alpha}, where the triplet loss again outperforms NT-Xent by a significant amount on both datasets. Notably, NT-Xent's performance is improved on real-world data by using a smaller $\tau$ parameter. Based on these results, we choose the triplet loss with $N_{\text{A}} = 512$, $N_{\text{PPA}}=2$, and $\alpha=0.5$ for our NMFP model.

\section{Results}\label{sec:results}

Here, we compare NMFP with NAFP-Adam (baseline method) and the only state-of-the-art models with publicly available weights: NAFP-LAMB~\cite{chang_neural_2021} and GraFP~\cite{bhattacharjee_grafprint_2025}. 
NMFP and NAFP models are trained at 8\,kHz sampling rate, while GraFP was trained at 16\,kHz, hence requiring upsampling. 
Additionally, GraFP uses a 100\,ms hop, whereas our models operate with 500\,ms. We consider GraFP with each hop duration.

Table~\ref{tab:bmat} shows that NMFP substantially outperforms both NAFP and GraFP on track-level identification. 
In particular, NMFP improves its baseline (NAFP-Adam) by 14.4\% on synthetic data and 30.3\% on real-world data, and it surpasses the official NAFP-LAMB by 12.8\% and 21.9\%, respectively. 
Against GraFP, NMFP scores 69.6\% higher on synthetic data and 43.9\% on real-world data when using a 500\,ms hop.
With a 100\,ms hop, it still outperforms GraFP by 66.9\% on synthetic data and 17.3\% on real-world data. 
This drastic difference could be due to the applied upsampling, which can not create the higher frequencies that GraFP likely depends on.

In Table~\ref{tab:segment}, we compare NMFP against NAFP on segment-level identification on the synthetic data. We exclude GraFP from this comparison, as its authors do not consider segment-level identification, and due to their peak picking methods' interaction with silent segments. 
NMFP outperforms its baseline, NAFP-Adam, by 13.3\% in exact matches and 14.4\% in near matches.  
NMFP also outperforms NAFP-LAMB by 8.7\% in exact matches and 12.9\% in near matches. Together, the results in Table~\ref{tab:bmat} and Table~\ref{tab:segment} demonstrate that our model, NMFP, sets the state-of-the-art on both track- and segment-level MI.

\begin{table}[t]
    \renewcommand{\arraystretch}{0.85}
    \centering
    \setlength{\tabcolsep}{5pt}
    \begin{tabular}{lccccc}
    \toprule
        \multirow{2}{*}{\vspace{-5pt}Model} & \multirow{2}{*}{\vspace{-5pt}Match} & \multicolumn{4}{c}{Top-1 hit rate (\%)} \\
        \cmidrule(lr){3-6}
         & & 1\,s & 2\,s & 5\,s & 10\,s \\
    \midrule
        \multirow{2}{*}{NAFP-Adam} & exact & 50.5 & 64.8 & 74.1 & 78.2 \\
         & near & 61.2 & 74.2 & 83.4 & 87.7 \\
         \cmidrule(lr){2-6}
        \multirow{2}{*}{NAFP-LAMB~\cite{chang_neural_2021}} & exact & 55.1 & 66.4 & 74.4 & 77.9 \\
         & near & 62.7 & 74.6 & 83.3 & 87.4 \\
    \midrule
            \multirow{2}{*}{NMFP (proposed)} & exact & \textbf{63.8} & \textbf{74.8} & \textbf{82.0} & \textbf{85.0} \\
             & near & \textbf{75.6} & \textbf{83.5} & \textbf{89.2} & \textbf{92.0} \\
    \bottomrule
    \end{tabular}
    \caption{
    Segment-level identification results on the synthetic dataset.
    }
    \label{tab:segment}
\end{table}

\section{Conclusion}\label{sec:conclusion}

We present a comprehensive framework for enhancing the robustness of neural AFP models against real‑world audio degradation. By correcting evaluation flaws in prior work, we establish a more reliable benchmark for future AFP research. Our evaluations, conducted on both a synthetic dataset and a real-world dataset recorded in diverse music venues, show that NMFP significantly outperforms existing neural AFP models with publicly available weights. 
Specifically, on track-level identification, it outperforms the official NAFP model by 12.9\% on synthetic data and by 21.9\% on real-world data. 

Our success stems from two key areas. First, we show that paying careful attention to musical signal properties and room acoustics enhances performance considerably. Second, by revisiting metric learning, we uncovered several key findings that further improve performance. We discovered that the triplet loss, despite common assumptions, outperforms modern alternatives such as NT-Xent. We also found that triplet loss does not suffer from the performance saturation seen with NT-Xent at large batch sizes. Finally, we characterized a critical trade‑off between the number of anchors and positives per anchor in training batches. Together, these insights form a set of validated, high-impact principles for neural AFP development.

\clearpage

\section{Acknowledgments}
This work was supported by the pre-doctoral program AGAUR-FI ajuts (2024 FI-3 00065) Joan Oró, funded by the Secretaria d’Universitats i Recerca of the Departament de Recerca i Universitats of the Generalitat de Catalunya; and by the Cátedras ENIA program “IA y Música: Cátedra en Inteligencia Artificial y Música” (TSI-100929-2023-1), funded by the Secretaría de Estado de Digitalización e Inteligencia Artificial and the European Union – Next Generation EU.

This work was also part of the project TROBA – Technologies for the recognition of musical works in the era of dynamic generation of audio content (ACE014/20/000051), within the call Nuclis d’R+D 2024, with the support of ACCIÓ (Agency for Business Competitiveness, Government of Catalonia).

\bibliography{Fingerprinting}

\begin{thebibliography}{10}
\providecommand{\url}[1]{#1}
\csname url@samestyle\endcsname
\providecommand{\newblock}{\relax}
\providecommand{\bibinfo}[2]{#2}
\providecommand{\BIBentrySTDinterwordspacing}{\spaceskip=0pt\relax}
\providecommand{\BIBentryALTinterwordstretchfactor}{4}
\providecommand{\BIBentryALTinterwordspacing}{\spaceskip=\fontdimen2\font plus
\BIBentryALTinterwordstretchfactor\fontdimen3\font minus \fontdimen4\font\relax}
\providecommand{\BIBforeignlanguage}[2]{{%
\expandafter\ifx\csname l@#1\endcsname\relax
\typeout{** WARNING: IEEEtran.bst: No hyphenation pattern has been}%
\typeout{** loaded for the language `#1'. Using the pattern for}%
\typeout{** the default language instead.}%
\else
\language=\csname l@#1\endcsname
\fi
#2}}
\providecommand{\BIBdecl}{\relax}
\BIBdecl

\bibitem{haitsma_robust_2001}
J.~Haitsma, T.~Kalker, and J.~Oostveen, ``Robust audio hashing for content identification,'' in \emph{Int. {Workshop} on {Content}-{Based} {Multimedia} {Indexing} ({CBMI})}, 2001.

\bibitem{haitsma_highly_2002}
J.~Haitsma and T.~Kalker, ``A highly robust audio fingerprinting system,'' in \emph{Proc. of the 3rd {Int}. {Conf}. on {Music} {Information} {Retrieval} ({ISMIR})}, 2002.

\bibitem{cano_review_2002}
P.~Cano, E.~Batle, T.~Kalker, and J.~Haitsma, ``A review of algorithms for audio fingerprinting,'' in \emph{{IEEE} {Workshop} on {Multimedia} {Signal} {Processing} ({MMSP})}, 2002.

\bibitem{wang_shazam_2006}
A.~Wang, ``The {Shazam} music recognition service,'' \emph{Communications of the ACM}, vol.~49, no.~8, pp. 44--48, 2006.

\bibitem{burges_distortion_2003}
C.~Burges, J.~Platt, and S.~Jana, ``Distortion discriminant analysis for audio fingerprinting,'' \emph{IEEE Transactions on Speech and Audio Processing}, vol.~11, no.~3, pp. 165--174, 2003.

\bibitem{cortes_baf_2022}
G.~Cortès, A.~Ciurana, E.~Molina, M.~Miron, O.~Meyers, J.~Six, and X.~Serra, ``{BAF}: {An} audio fingerprinting dataset for broadcast monitoring,'' in \emph{Proc. of the 23rd {Int}. {Soc}. for {Music} {Information} {Retrieval} {Conf}. ({ISMIR})}, 2022.

\bibitem{arcas_now_2017}
\BIBentryALTinterwordspacing
B.~A.~y. Arcas, B.~Gfeller, R.~Guo, K.~Kilgour, S.~Kumar, J.~Lyon, J.~Odell, M.~Ritter, D.~Roblek, M.~Sharifi, and M.~Velimirović, ``Now {Playing}: {Continuous} low-power music recognition,'' 2017, arXiv:1711.10958 [cs, eess]. [Online]. Available: \url{http://arxiv.org/abs/1711.10958}
\BIBentrySTDinterwordspacing

\bibitem{yu_contrastive_2020}
\BIBentryALTinterwordspacing
Z.~Yu, X.~Du, B.~Zhu, and Z.~Ma, ``Contrastive unsupervised learning for audio fingerprinting,'' 2020, arXiv:2010.13540 [cs, eess]. [Online]. Available: \url{http://arxiv.org/abs/2010.13540}
\BIBentrySTDinterwordspacing

\bibitem{chang_neural_2021}
S.~Chang, D.~Lee, J.~Park, H.~Lim, K.~Lee, K.~Ko, and Y.~Han, ``Neural audio fingerprint for high-specific audio retrieval based on contrastive learning,'' in \emph{{IEEE} {Int}. {Conf}. on {Acoustics}, {Speech} and {Signal} {Processing} ({ICASSP})}, 2021.

\bibitem{singh_attention-based_2022}
A.~Singh, K.~Demuynck, and V.~Arora, ``Attention-based audio embeddings for query-by-example,'' in \emph{Proc. of the 23rd {Int}. {Society} for {Music} {Information} {Retrieval} {Conf}. ({ISMIR})}, 2022.

\bibitem{bhattacharjee_grafprint_2025}
A.~Bhattacharjee, S.~Singh, and E.~Benetos, ``{GraFPrint}: {A} {GNN}-based approach for audio identification,'' in \emph{{IEEE} {Int}. {Conf}. on {Acoustics}, {Speech} and {Signal} {Processing} ({ICASSP})}, 2025.

\bibitem{chen_simple_2020}
T.~Chen, S.~Kornblith, M.~Norouzi, and G.~Hinton, ``A simple framework for contrastive learning of visual representations,'' in \emph{Proc. of the 37th {Int}. {Conf}. on {Machine} {Learning} ({ICML})}, 2020.

\bibitem{serra_supervised_2025}
J.~Serrà, R.~O. Araz, D.~Bogdanov, and Y.~Mitsufuji, ``Supervised contrastive learning from weakly-labeled audio segments for musical version matching,'' in \emph{Proc. of the 42nd {Int}. {Conf}. on {Machine} {Learning} ({ICML})}, 2025.

\bibitem{guinot_semi-supervised_2024}
J.~Guinot, E.~Quinton, and G.~Fazekas, ``Semi-supervised contrastive learning of musical representations,'' in \emph{Proc. of the 25th {Int}. {Soc}. for {Music} {Information} {Retrieval} {Conf}. ({ISMIR})}, 2024.

\bibitem{schroff_facenet_2015}
F.~Schroff, D.~Kalenichenko, and J.~Philbin, ``Facenet: {A} unified embedding for face recognition and clustering,'' in \emph{{IEEE} {Conf}. on {Computer} {Vision} and {Pattern} {Recognition} ({CVPR})}, 2015, pp. 815--823.

\bibitem{yeh_decoupled_2022}
C.-H. Yeh, C.-Y. Hong, Y.-C. Hsu, T.-L. Liu, Y.~Chen, and Y.~LeCun, ``Decoupled contrastive learning,'' in \emph{Computer {Vision} -- {ECCV}}, 2022.

\bibitem{wang_understanding_2020}
T.~Wang and P.~Isola, ``Understanding contrastive representation learning through alignment and uniformity on the hypersphere,'' in \emph{Proc. of the 37th {Int}. {Conf}. on {Machine} {Learning} ({ICML})}, 2020.

\bibitem{koromilas_bridging_2024}
P.~Koromilas, G.~Bouritsas, T.~Giannakopoulos, M.~A. Nicolaou, and Y.~Panagakis, ``Bridging mini-batch and asymptotic analysis in contrastive learning: {From} {InfoNCE} to kernel-based losses,'' in \emph{Proc. of the 41st {Int}. {Conf}. on {Machine} {Learning} ({ICML})}, 2024.

\bibitem{khosla_supervised_2020}
P.~Khosla, P.~Teterwak, C.~Wang, A.~Sarna, Y.~Tian, P.~Isola, A.~Maschinot, C.~Liu, and D.~Krishnan, ``Supervised contrastive learning,'' in \emph{Proc. of the 34th {Int}. {Conf}. on {Neural} {Information} {Processing} {Systems} ({NeurIPS})}, 2020.

\bibitem{tian_stablerep_2023}
Y.~Tian, L.~Fan, P.~Isola, H.~Chang, and D.~Krishnan, ``{StableRep}: {Synthetic} images from text-to-image models make strong visual representation learners,'' in \emph{Proc. of the 37th {Int}. {Conf}. on {Neural} {Information} {Processing} {Systems} ({NeurIPS})}, 2023.

\bibitem{defferrard_fma_2017}
M.~Defferrard, K.~Benzi, P.~Vandergheynst, and X.~Bresson, ``{FMA}: {A} dataset for music analysis,'' in \emph{Proc. of the 18th {Int}. {Soc}. for {Music} {Information} {Retrieval} {Conf}. ({ISMIR})}, 2017.

\bibitem{mesaros_tut_2016}
A.~Mesaros, T.~Heittola, and T.~Virtanen, ``{TUT} database for acoustic scene classification and sound event detection,'' in \emph{24th {European} {Signal} {Processing} {Conf}. ({EUSIPCO})}, 2016.

\bibitem{murphy_openair_2010}
D.~T. Murphy and S.~Shelley, ``{OpenAIR}: {An} {Interactive} {Auralization} {Web} {Resource} and {Database},'' in \emph{Audio {Engineering} {Society} {Convention} 129}, 2010.

\bibitem{jeub_binaural_2009}
M.~Jeub, M.~Schäfer, and P.~Vary, ``A binaural room impulse response database for the evaluation of dereverberation algorithms,'' in \emph{Proc. of {Int}. {Conf}. on {Digital} {Signal} {Processing} ({DSP})}, 2009.

\bibitem{traer_statistics_2016}
J.~Traer and J.~H. McDermott, ``Statistics of natural reverberation enable perceptual separation of sound and space,'' \emph{Proc. of the National Academy of Sciences}, vol. 113, no.~48, pp. E7856--E7865, 2016.

\bibitem{franco_multi-angle_2022}
J.~Franco, B.~Bǎcilǎ, T.~Brookes, and E.~De~Sena, ``A multi-angle, multi-distance dataset of microphone impulse responses,'' \emph{Journal of the Audio Engineering Society}, 2022.

\bibitem{kingma_adam_2015}
D.~Kingma and J.~Ba, ``Adam: {A} method for stochastic optimization,'' in \emph{Proc. of the 3rd {Int}. {Conf}. for {Learning} {Representations} ({ICLR})}, 2015.

\bibitem{you_large_2020}
Y.~You, J.~Li, S.~Reddi, J.~Hseu, S.~Kumar, S.~Bhojanapalli, X.~Song, J.~Demmel, K.~Keutzer, and C.-J. Hsieh, ``Large batch optimization for deep learning: {Training} {BERT} in 76 minutes,'' in \emph{Proc. of the 8th {Int}. {Conf}. on {Learning} {Representations} ({ICLR})}, 2020.

\bibitem{park_specaugment_2019}
D.~S. Park, W.~Chan, Y.~Zhang, C.-C. Chiu, B.~Zoph, E.~D. Cubuk, and Q.~V. Le, ``{SpecAugment}: {A} simple data augmentation method for automatic speech recognition,'' in \emph{20th {Annual} {Conf}. of the {Int}. {Speech} {Communication} {Association} ({INTERSPEECH})}, 2019.

\bibitem{johnson_billion-scale_2019}
J.~Johnson, M.~Douze, and H.~Jégou, ``Billion-scale similarity search with {GPUs},'' \emph{IEEE Transactions on Big Data}, vol.~7, no.~3, pp. 535--547, 2019.

\bibitem{sonnleitner_robust_2016}
R.~Sonnleitner and G.~Widmer, ``Robust quad-based audio fingerprinting,'' \emph{IEEE/ACM Trans. on Audio, Speech, and Language Processing}, vol.~24, no.~3, pp. 409--421, 2016.

\bibitem{baez-suarez_samaf_2020}
A.~Báez-Suárez, N.~Shah, J.~A. Nolazco-Flores, S.-H.~S. Huang, O.~Gnawali, and W.~Shi, ``{SAMAF}: {Sequence}-to-sequence autoencoder model for audio fingerprinting,'' \emph{ACM Trans. Multimedia Comput. Commun. Appl.}, vol.~16, no.~2, pp. 1--23, 2020.

\bibitem{agarwaal_robust_2023}
\BIBentryALTinterwordspacing
A.~Agarwaal, P.~Kanaujia, S.~S. Roy, and S.~Ghose, ``Robust and lightweight audio fingerprint for automatic content recognition,'' 2023, arXiv:2305.09559 [cs, eess]. [Online]. Available: \url{http://arxiv.org/abs/2305.09559}
\BIBentrySTDinterwordspacing

\bibitem{ramona_audioprint_2013}
M.~Ramona and G.~Peeters, ``{AudioPrint}: {An} efficient audio fingerprint system based on a novel cost-less synchronization scheme,'' in \emph{{IEEE} {Int}. {Conf}. on {Acoustics}, {Speech} and {Signal} {Processing} ({ICASSP})}, 2013.

\bibitem{beranek_concert_1992}
L.~L. Beranek, ``Concert hall acoustics—1992,'' \emph{The Journal of the Acoustical Society of America}, vol.~92, no.~1, pp. 1--39, 1992.

\end{thebibliography}

\end{document}